# On electromagnetic induction in electric conductors

ALEXANDER I. KOROLEV

Department of General Physics I, Faculty of Physics, St. Petersburg State University, 7/9 Universitetskaya nab., 199034, Saint Petersburg, Russian Federation

e-mail: alex-korolev@ya.ru

phone: +79045517888

Experimental validation of the Faraday's law of electromagnetic induction (EMI) is performed when an electromotive force is generated in thin copper turns, located inside a large magnetic coil. It has been established that the electromotive force (emf) value should be dependent not only on changes of the magnetic induction flux through a turn and on symmetry of its crossing by magnetic power lines also. The law of EMI is applicable in sufficient approximation in case of the changes of the magnetic field near the turn are symmetrical. Experimental study of the induced emf in arcs and a direct section of the conductor placed into the variable field has been carried out. Linear dependence of the induced emf on the length of the arc has been ascertained in case of the magnetic field distribution symmetry about it. Influence of the magnetic field symmetry on the induced emf in the arc has been observed. The curve of the induced emf in the direct section over period of current pulse is similar to this one for the turns and arcs. The general law of EMI for a curvilinear conductor has been deduced. Calculation of the induced emf in the turns wrapped over it and comparison with the experimental data has been made. The proportionality factor has been ascertained for the law. Special conditions have been described, when the induced emf may not exist in the presence of inductive current. Theoretical estimation of the inductive current has been made at a induced low voltage in the turn. It has been noted the necessity to take into account the concentration of current carriers in calculation of the induced emf in semiconductors and ionized conductors.

# Introduction

For the first time the phenomenon of induction of electrical current in a conductor under a variable magnetic field was described by F. Zantedeschied in 1829 and by M. Faraday in 1831 [1]. M. Faraday established that "electrotonic" state in a conductor appears at crossing by magnetic power lines. Many experiments on quantitative investigation of "magnetoelectric" induction were carried out. A description that allows detecting the induction current direction knowing the mode of magnetic power lines movement was given. A brief metaphysical rule of the direction detecting was formulated by E. Lenz in 1832 [2]. First mathematical expression of the emi law was presented by F. Neumann in 1845 -1847 [3]. He introduced a concept of "vector - potential ", expressing it through the induction of magnetic field. Contribution to the theory of electromagnetism was made by Felici.

Wide theoretical description of electromagnetic phenomena was made by J. Maxwell in the middle of 19th century [4,5]. The scientist put the experimental results, obtained by M. Faraday, into the language of mathematics. The expression for components of the emf in a loop was given as a system of three scalar equations (1).

$$\begin{cases} P = -\dfrac{dF}{dt} - \dfrac{d\psi}{dx} \\ Q = -\dfrac{dG}{dt} - \dfrac{d\psi}{dy} \\ R = -\dfrac{dH}{dt} - \dfrac{d\psi}{dz} \end{cases} \quad (1)$$

Here P, Q and R are the components of the emf representing the potential drop per unit of length of a conductor along x, y, z axes of Cartesian coordinates, respectively, at t point of time. F, G and H are projections of the electromagnetic momentum on x, y, z axes. ψ is the electrical potential at the point under consideration. Expression for an induced emf in a fixed loop of arbitrary shape is

$$\xi = \int \left( P\frac{dx}{ds} + Q\frac{dy}{ds} + R\frac{dz}{ds} \right) ds \quad (2)$$

Where dS is an element of the contour of integration. At integration, the terms with ψ are cancelled. Thus, ξ is defined by the derivatives of the electromagnetic momentum. This is the result of application of the second Newton's law to the current carriers in the loop.

Further contribution to electromagnetic theory was made by O. Heaviside [6]. He transformed some Maxwell's equations using the terms of three-dimensional vectors, intensities and induction of electrical and magnetic fields. The discovery of electromagnetic waves by Hertz promoted the deduction of two well- known equations of Hertz- Heaviside (now they are called as rotary Maxwell's equations). There the electric and magnetic fields intensities are related, irrespective of the presence of particles in the medium that let us to describe the propagation of electromagnetic waves in some approximation. Nevertheless, it does not mean that one variable field induces another one [7,8]. In [7] there is a conclusion that the momentum in (1) is induced by "outside forces" of the variable magnetic field. The expression for these forces via vector-potential is given in [8].

In classical electrodynamics the law of emi for a closed loop (3) is an integral equivalent for the first Hertz- Heaviside's equation for vacuum (4) [9, 10].

$$\varepsilon_i = -\frac{\partial \Phi}{\partial t} \quad (3)$$

$$\nabla \times \boldsymbol{E} = -\frac{\partial \boldsymbol{B}}{\partial t} \quad (4)$$

According to (3), the induced emf $\varepsilon_i$ is expressed by a time derivative of the magnetic flux $\Phi$ through a loop (in SI). However, the electric and magnetic fields in a real conductor differ from these in transverse electromagnetic wave in vacuum and they depend on the conductor charges and its material. As well as it is shown in [11] the variable magnetic field can propagate independently of the variable electrical one. Thus, the transfer from (4) to (3) is not correct. And the paradoxes are known when calculation of the induced emf using (3) without taking additional reservations [12,13] into account are made. In this connection, the experimental verification of (3) and formulation of more exact law for emi for a conductor of arbitrary form is of our interest.

## Measurements of magnetic field in the coil using induction sensors

The magnetic coil is a source of the field (see Fig. 1) is winded by a copper bus in textile isolation with section of 2×4 mm$^2$. Number of turns is 40, the inner diameter of winding is 12 mm, the outer diameter is 35 mm, the length - 40 mm. The winding of the coil enables to make current outlets far from its axis. A 3 mm textolite lining is inserted into the central part of the coil to avoid the electrical breakdown (see Fig. 1). The coil is placed inside a cast- iron bandage and tightened with thick isolating jaws on each side. The coil is intended for generation of a strong magnetic field but is suitable for generation of weak fields too.

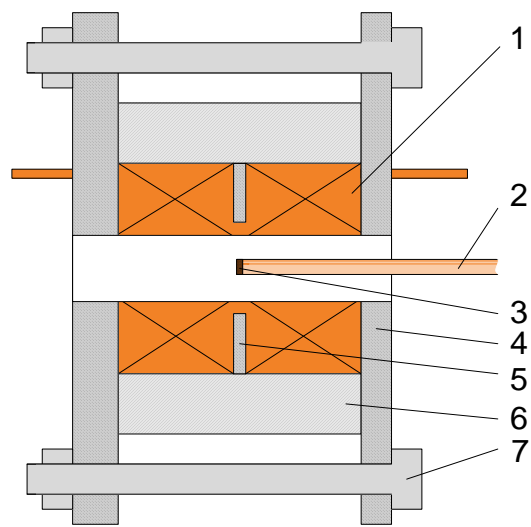

*Fig. 1. Axial section of the magnetic coil. 1 - coil windings, 2 - induction sensor, 3 - sensor winding, 4 - isolating jaw, 5 - isolating lining, 6 - cast- iron bandage, 7 -bolt .*

Induction sensors (2 pcs.) are thin ceramic tubes 2.5 mm and 5.5 mm in diameter with copper windings at the ends (see Fig. 1). The wire is covered by varnish isolation, the wire diameter is 0.1 mm, the number of turns is 10. The turns are tight to each other forming coils 1 mm in length. The current outlets of the coils are parallel to the tubes. The sensors are fixed at a support so that their axes coincide with the axe of the magnetic coil. They are placed into the coil in this position. The Hall sensor Honeywell SS496A is used to measure magnetic fields up to 0.4 T. Time response of the sensor is 3 µs, average sensitivity under normal climatic conditions is 2.4 mV/G.

Electrical scheme of the setup is presented in Fig. 2. Electrical energy is stored up in the capacitor bank, with capacity 2850 µF and maximal voltage 5 kV, and then it is directed to the coil through the mercury switch. The switch is closing when a triggering pulse is fed from the G5- 56 pulse generator. Pulses from the magnetic field sensors are tested with the C1- 73 oscilloscope. Photographs of the pulses are registered by the camera and then processed on PC. Because of great attenuation the first current pulse is only important. Value of the second (reverse) pulse is no more than 17% of the first one. Next pulses are not taken into account.

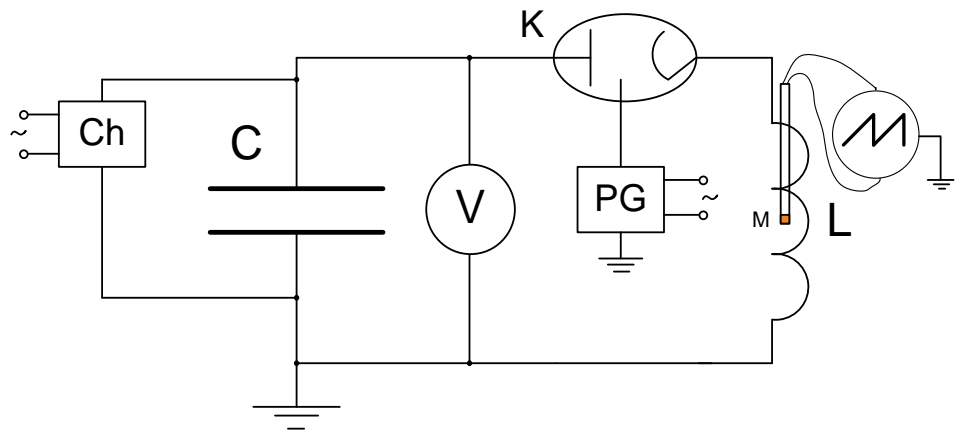

*Fig. 2.Scheme of the experimental setup. C- capacitor bank, L- magnetic coil, M- induction sensor, K- mercury switch, PG- pulse generator, Ch- charger for capacitor bank.*

Dependences of magnetic field on time are obtained in two ways. The first way to define *B(t)* is to measure and process a signal from the Hall sensor (fields are less than 0.4 T). The second one is to calculate the magnetic field by signals from the induction sensors using formula (3). An expression for module of magnetic induction in the area of the sensor at *t* point of time is

$$B(t) = \frac{4}{\pi d^2 N} \int_0^t \varepsilon_i dt \quad (5)$$

There d is the diameter of the sensor winding, N= 10 is the number of turns, $\varepsilon_i$ is the measured induced emf.

The normalized curves of magnetic induction in the center of coil are presented in Fig. 3. The normalization is made to maximum magnetic induction. To make it easy the integration is performed up to a peak of the second pulse. This makes additions in time and values of B up to 20 %.

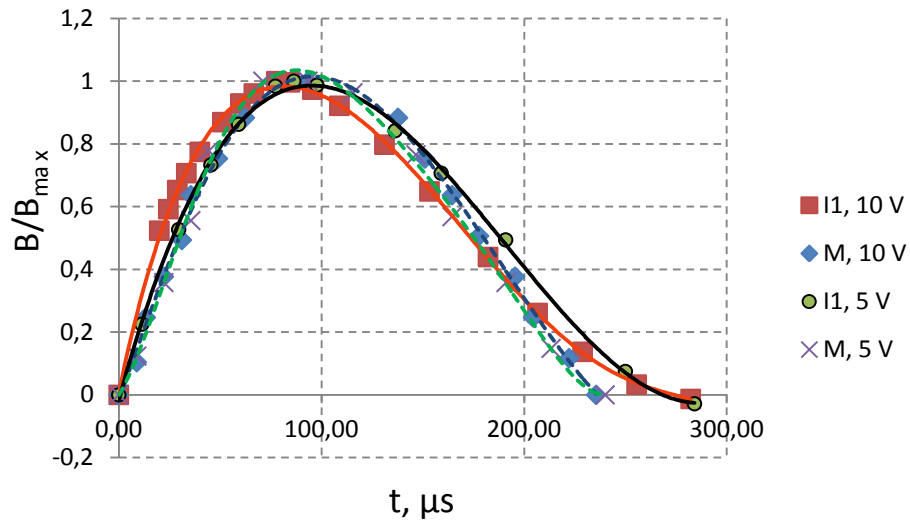

Fig. 3.The curves of normalized magnetic induction in the center of coil during the first pulse. I1- induction sensor with diameter 5.5 mm. M- Hall sensor. Voltages of charged capacitor bank are shown.

Peak values of the magnetic induction in the center and on face of the coil are given in table 1.

| Values $B_{max}$, T | Sensor 2,5 mm | Sensor 5,5 mm | Hall sensor |
|---|---|---|---|
| center, $U_0$ = 10 V | 0,018 | 0,017 | 0,034 |
| center, $U_0$ = 5 V |  | 0,009 | 0,014 |
| flank, $U_0$ = 10 V | 0,010 | 0,011 | 0,017 |
| flank, $U_0$ = 5 V |  | 0,007 | 0,007 |

Table 1. Peak values of the magnetic inductions calculated using induction sensors and Hall sensor in the center and on face of the coil along its axe. $U_0$ - the voltage of charged capacitor bank.

One can see that the shapes of curves are in close agreement. The values of magnetic induction calculated with (5) are 1-2 times less than these measured using the Hall sensor. The calculated $B_{max}$ values are the same for both induction sensors within the error. The differences between $B_{max}$ values calculated for the sensors in the coil center are more than these ones at the face. The error in B values is connected with the field homogeneity and accuracy of the sensor positions, besides the error of calculation. The dimensions of Hall sensor are 4×6 mm so it is impossible to investigate the field homogeneity.

Thus, we can conclude that the induced emf in the sensors being used is directly proportional to the rate of change of magnetic induction: $\varepsilon_i \sim \frac{\partial B}{\partial t}$ , and the difference in values of emf does nor exceed a permissible error of the measurement. So, the emi law in form of (3) is applicable to calculation of the induced emf in sensors placed axially inside the magnetic coil. It is in agreement with the theory of induction coil sensors [14].

## Measurement of induced emf in open-loop curvilinear conductors

Emf is induced invariably when magnetic power lines cross a conductor triggering changes of the magnetic field. That was noted by M. Faraday in 1832. However, the law (3) describes the only class of closed conductors. Emf induced in a curvilinear conductor may be caused by a variable magnetic field or by movement of a conductor in the area of heterogeneous magnetic field. A special case is the so- called unipolar induction, which is caused by movement of electrons in a conductor moving in homogeneous magnetic field. Its mechanism is determined by Lorentz forces acting upon the moving electrons.

Let us investigate the generation of induced emf in conductors of curvilinear form. Any spatial curve with non- zero curvature may be approximated by a set of circlar arcs. Thus, the problem of determination of induced emf in a curvilinear conductor is to find a value and direction of induced emf in an arc. The parameters of arc are length, radius and position relative to magnetic power lines.

### *Induced emf in a wire arc of finite curvature*

The arcs are made of copper wire with 0.1 mm in diameter and are fixed on a dielectric cylinder in the center of magnetic coil, normal to its axe (see Fig. 1). The ends of the arcs are bended so that the current outlets are parallel to the axe (see Fig. 4).

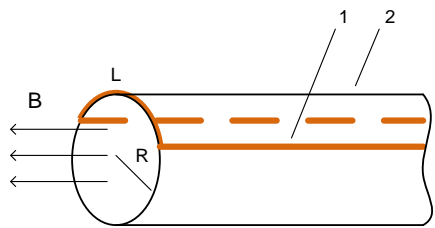

*Fig. 4. Shceme of a wire arc 1 fixed on a thin dielectric cylinder 2. L is a length, R=2.75 mm is a radius.*

The curves of emf induced in the arcs with different length and fixed radius of 2.75 mm are presented in Fig. 5. The time is limited by a peak of the second pulse, when derivative of *B(t)* vanishes. Minimums of the curves correspond to the inflections of dependencies *B(t)* obtained using Hall sensor.

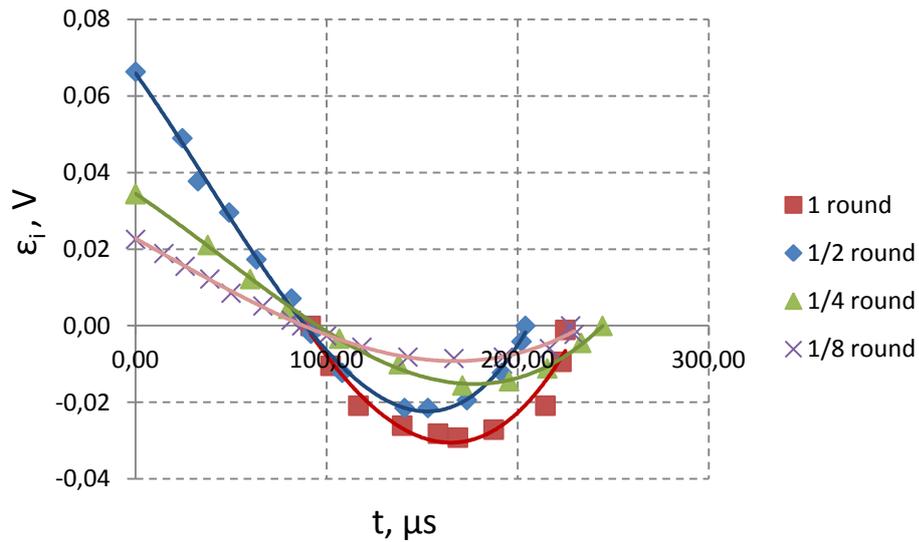

Fig. 5. Induced EMF in the arcs (pieces of the turn). Length of a turn is 17,3 mm. The voltage of charged capacitor bank is 100V.

One can see the curves differ by its stretch along the vertical axis and by slight shift along the horizontal one. To compare emf let us take the minimum points. Relationships between the emf absolute values in minimums and the arc length are presented in Fig. 6. Graph (a) is made from the data given in Fig. 5, graph (b) from the average values of induced emf: for a whole turn (2 measurings), half turn (4 measurings), 1/8 turn (3 measurings). In measuring, the arcs are rotated in the plane normal to the coil axis. The relations may be approximated by straight lines. Spread of points arises from inaccuracy of arc positions in the coil center and heterogeneity of the magnetic field inside the coil. Graph (a) does not pass through the coordinate origin because of the large spread of emf values in small arcs (up to 5 times for the arcs 1/8 turn in length). Graph (b) is more exact, because it is obtained using average data. From the physical point of view it is clear that induced emf is to be equal to zero in an arc of zero length.

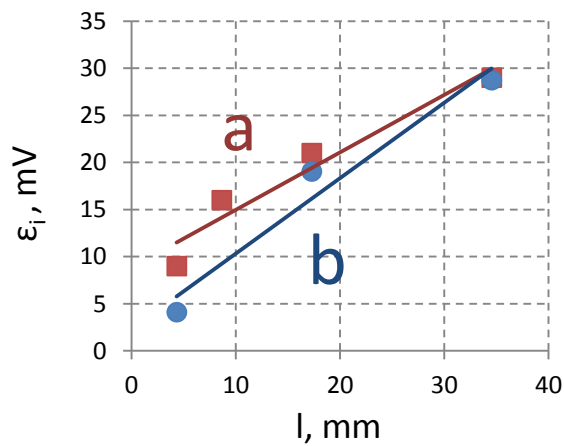

Fig. 6. Dependencies of absolute values of induced emf in arcs on their lengths. a - using data presented at fig. 5, b - average values. The arcs are placed in the coil center, in the plane normal to the coil axis. Radius of the arcs is 2.75 mm.

The family of curves of emf is obtained similarly for the arcs with different radii and of fixed length 5.5 mm. But the resulting relationships are ambiguous. That is because of the value of induced emf depends on the arc position in the normal plane. The emf decreases considerably when the center of gravity of the arc is shifted to the coil center. The emf is lacking when a wire segment ($R_c = \infty$) is placed symmetrically in the center of coil. Power lines of the variable magnetic field do not cross the segment. Small induced EMF is observed when the segment is shifting from the center of the coil.

Thus, great spread of points of $\varepsilon_i(l)$ and differences in calculated values of magnetic induction in the section above can be explained as follows. Shift of the inductive sensors in the normal plane from the center causes low values of the magnetic induction. Shift of the arcs from the center leads to increase of the induced emf. At that the summary magnetic induction changes insignificantly, but its components in different pieces of the coil winding change essentially.

The proportionality of $\varepsilon_i \sim l$ with fixed $R_c$ is in agreement with proportionality of $\varepsilon_i$ to the area of the sector of inscribed circle. However, $\varepsilon_i$ must be proportional to the circular segment area (the flux through other part of the loop is negligible). This contraries to proportionality to the length and makes us to revise the law (3).

*Induced EMF in a straight wire*

The experimental setup (Fig. 2) is rebuilt to investigate the induced emf in a straight wire. A ferrite rod is inserted into the magnetic coil. The rod is 9 mm in diameter, the length of its open part is 140 mm. A segment of copper wire 400 mm long and 0.1 mm in diameter is placed perpendicularly at a distance of 100 mm from the end of the rod (see Fig. 7). Minimal distance from the wire to the rod is 2 mm. Voltage in the wire is registered by an oscilloscope.

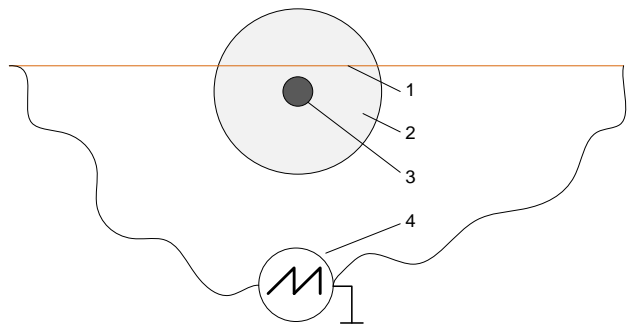

*Fig. 7. Scheme of measuring of induced emf in a straight wire. 1 is a segment of thin wire 400 mm long, 2 is a magnetic coil, 3 is a ferrite rod, 4 is an oscilloscope.*

The dependence of induced emf on time is shown in Fig. 8. The voltage of charged capacitor bank is 100 V. One can see that the shape of the curve during the first pulse (200 μs) is similar to the curves for windings and arcs. Some stretching of the curve and bumps after the pulse can be explained by the ferrite retentivity.

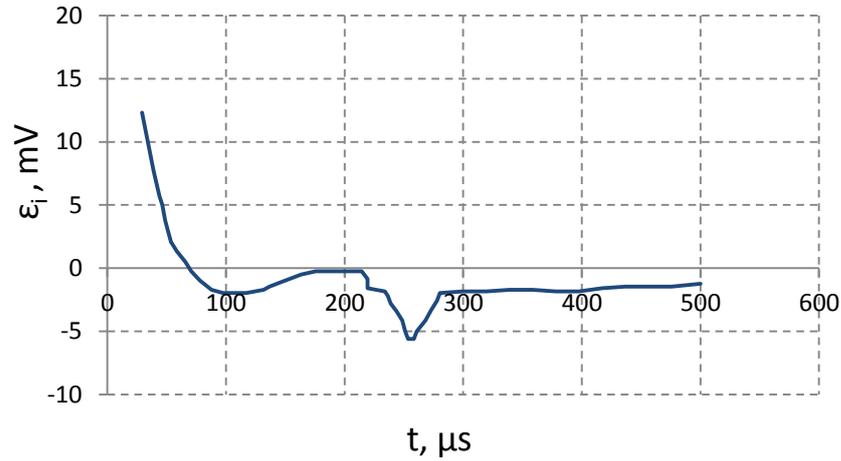

*Fig. 8. Induced EMF in a straight wire placed in the area of variable magnetic field near a ferrite rod.*

## The general law of EMI for curvilinear conductor

Two reasons of induced emf should be considered to formulate the general law of EMI for a curvilinear conductor. The first one is the magnetic force acting at free charges in the conductor that have average velocities close to zero. Presence of a variable magnetic field and asymmetry of its distribution near a conductor (crossing by magnetic power lines) are necessary for origin of this force. Let us suppose the direction of magnetic force is determined, by analogy with Lenz force, by the vector product of relative velocity of the charge across magnetic power lines and the vector of magnetic induction. Direction of a normal component in relation to the magnetic power lines is important. This assumption is in agreement with obtained experimental data and Lenz's rule. Direction of the velocity at any moment can be defined by a gradient and a sign of derivative of the magnetic induction at the point of the charge. The magnetic force, acting at the charge in some section of the conductor from the magnetic dipole placed at a spatial point with coordinates x, y, z (see fig. 9), is:

$$\vec{F_M} \sim q \left[ \vec{e_{xyz}} \times \frac{\partial \vec{B_{xyz}}}{\partial t} \right] \quad (6)$$

Here q is the charge, $\vec{e_{xyz}}$ is the unit vector directed along the gradient of the magnetic induction, $B_{xyz}$ is the magnetic induction produced by the dipole at the point of the charge location. Vector $\frac{\partial \vec{B_{xyz}}}{\partial t}$ is collinear to vector $\vec{B_{xyz}}$ at every point of time.

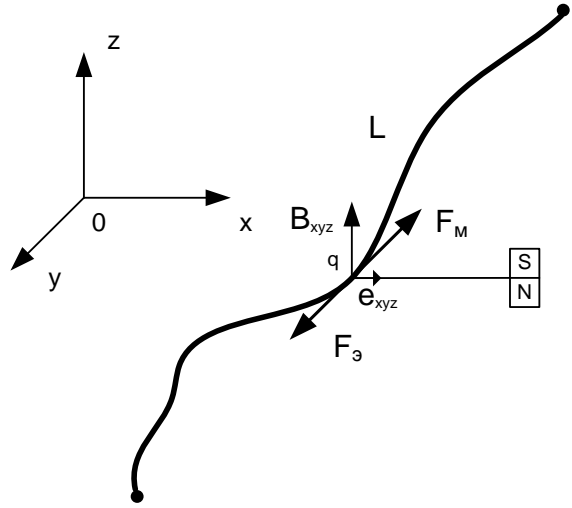

*Fig. 9. Illustration of origin of magnetic forces acting at charge q in a conductor L by variable magnetic field of the dipole. Directions of the forces are shown for increasing magnetic field, q < 0.*

Summary magnetic force of all magnetic dipoles in the neighborhood *V* of the conductor is: $\overrightarrow{F_{M\Sigma}} = \sum_V \overrightarrow{F_M}$. Shift of the charges induces an electrical field. The shift is sufficiently lower when small currents flow in the conductor with high concentration of free charges. Like microscopic shifts causes stress in mechanics, the shift of charges causes electrical voltage. Thus, in case of zero current $F_{M\Sigma}$ is compensated by electrostatic force of the electrical field induced in the conductor:

$$\overrightarrow{F_{M\Sigma}} = -\vec{F_e} = -q\vec{E} \quad (7)$$

Let us obtain the required induced emf (voltage) integrating (7) along the line of conductor *L*:

$$\varepsilon_i(t) = -\frac{1}{q}\int_L (\overrightarrow{F_{M\Sigma}}, \overrightarrow{dl}) \sim -\int_L (\sum_V \left[\overrightarrow{e_{xyz}} \times \frac{\overrightarrow{\partial B_{xyz}}}{\partial t}\right], \overrightarrow{dl}) \quad (8)$$

As it follows from (8) the induced emf is proportional to the conductor length and to the sum of derivatives of the magnetic induction $B_{xyz}$ of dipoles (their sets) at every point of the conductor taking into account the orientation of the dipoles with respect to the conductor line. Proportionality to the square in the law (3) means proportionality to the length and radius of the turns. The proportionality to the length and derivatives in (8) is in agreement with (3). The proportionality to the radius is evident from the following considerations. Asymmetry of effect of the magnetic forces on the electrons in turns from the magnetic coil winding depends on distances between the segments of the turns and the coil axis (or the winding). According to Biot–Savart law, magnetic induction of the segment with current is inversely proportional to the square distance from it. The sum of magnetic inductions (and their derivatives) of the segments of the coil winding is proportional to the distance from the axis in sufficient approximation. This dependence is provided for in the vector sum of magnetic forces in (8). Verification of linearity of the dependence (squared dependence on the radius of a turn), calculation of the induced emf in the turns of inductive sensors, and determination of the proportionality constant in (8) are presented in the following section.

When a current is induced under variable magnetic field, the total voltage of induction decreases due to the effect of self-induction. It should also be noted that the induced emf is the demonstration of the phenomenon of EMI or with heterogeneities in the conductor structure or the irregularity of influence of the magnetic forces along the circuit investigated. If the whole circuit is placed in the area of a variable magnetic field and the potential drop is measured at its points, the last one depends on a degree of uniformity of the field and distribution of resistivity along the circuit. In case of total uniformity (for example, a closed loop with full symmetry of magnetic field close to it) the potential drop is lack. The magnetic forces will only generate induction currents in the turn. In this case the characteristic of the EMI effect is the magnitude of inductive current multiplied by electric resistance.

Let us make an estimation of the magnitude of inductive current in a circuit with low voltages. Let us use the provision of Drude's model, which describes the movement of electron gas in a conductor [15]. When moving free electrons experience viscous friction from atoms of the conductor. One can apply the second Newton's law to the current carriers, just as Maxwell's method (1), but taking into account the viscous friction and expression obtained for summary magnetic force of the variable magnetic field. So, the expression of movement of a current carrier in the conductor is:

$$m\frac{\overrightarrow{dv(t)}}{dt} = \overrightarrow{F_{M\Sigma}(t)} - \alpha\overrightarrow{v(t)} \quad (9)$$

Here m is the mass of the current carrier, v(t) is its velocity at *t* point of time, α is the friction coefficient (which is proportionate to conductivity of the conductor). Solving the differential equation (9) in projection on tangent to the conductor for some heterogeneity function (the magnetic force), we obtain the expression for v(t). Let us place it into the well- known expression for current density in the conductor and get an estimation of the inductive current:

$$j(t) = nqv(t) \quad (10)$$

There n is concentration of the current carriers in the conductor.

The second reason of the EMI phenomenon is Lorentz forces, acting on moving charges in a magnetic field. Unlike the magnetic forces described above, Lorentz forces do not move the charges but curve their paths. In this case the phenomenon of EMI is observed in so- called "unipolar" generators. The first generator of unipolar induction was the disk of Faradey- Arago [1], producing voltage of a few mV at great dimensions. However, the main mechanism of generation of the induced emf in the disc is the mechanism described above. That is because when the disk rotates the electrons experience an influence of the variable magnetic field (the magnet is near the disk edge). One of the first unipolar generators of EMI only due to Lorentz forces is the second experimental setup of Das Gupt [16]. Emf is generated in a conductive disk, placed coaxial with a disk magnet, when the disk rotates singly or jointly with the magnet. The induced emf in the unipolar generators is significantly less than the emf generated under magnetic forces (6) in coils. However, the unipolar generators produce a high current that limits their use in engineering [16]. Thus, the induced emf with mechanism of Lorentz forces makes a small correction in the induced EMF for a curvilinear conductor. We have the expression for the induced emf owing to Lorentz forces by analogy with derivation of (8):

$$\varepsilon_{iL} \sim - \int_L ([\vec{v} \times \vec{B}], \vec{dl}) \quad (11)$$

Here *v* is the velocity of charges in the laboratory reference system, in the section of conductor with coordinate *l*; *B* is the magnetic induction near the charge. Integration is performed along the conductor line. Summary induced emf for a curvilinear conductor with no current is:

$$E_i = \varepsilon_i + \varepsilon_{iL} \quad (12)$$

In the presence of current, the self- inducted emf should be considered, that reduces $E_i$..When the whole circuit is placed in the area of variable magnetic field, $E_i$ would be defined by the similarity of the picture of magnetic field along the circuit.

## Calculation of induced emf in windings using general law of EMI for curvilinear conductor

Let us make a calculation to verify the law (8) and to establish quantitative correspondence with the measured values of the induced emf in the induction sensors and wire turns. We consider a sensor placed in the center of magnetic coil athwart to its axis (see Fig. 10). The sensor winding consists of thin wire turns 1 mm in length that is significantly less than the length of the coil winding (36 mm). Therefore, distinction of the magnetic field near the turns is not significant. Thus, calculation of the induced emf may be done for one turn, placed in the center of the coil. Multiplying the obtained emf by the number of the turns in the sensor, we have the summary induced voltage, measured by the oscilloscope.

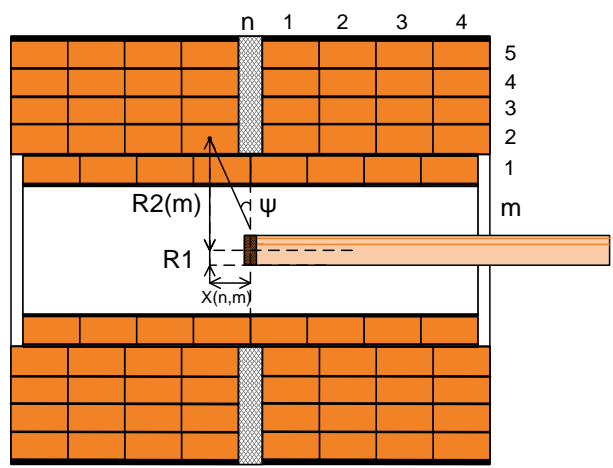

*Fig. 10. Axial section of magnetic coil winding and induction sensor.*

The emf in a turn of the sensor is induced by the variable magnetic field of currents flowing along the turns of the magnetic coil. Let us number the turns n = 1, 2, 3, 4, counting off along the axis from the centre; m = 1, 2, 3, 4, 5 is the number of a turn counted off athwart from the axis. In view of symmetry of the task, we can consider only half of the turns, placed on one of the sides of the plane of the sensor turn. We introduce the following notations: R1 is the radius of the sensor turn; R2(m) is the radius of a turn of the coil with number m; x(m,n) is the distance from the coil center to the center of the coil turn with numbers (m,n) on the axis; ψ is the angle between the plane of the sensor turn and direction to the center of section of some turn of the magnetic coil. Summing the emf from all turns of the coil, we obtain the emf in a turn of the sensor.

The explanatory illustration for calculation of the induced emf in a turn of the sensor is given in Fig. 11. Let us partition the source of magnetic field (the coil) into current elements (the sets of magnetic dipoles inside them). The magnetic induction $\vec{B}$ of the field, produced by an element of the current $\vec{dL}$, at the points of the sensor turn may is defined using Biot–Savart law. Let us determine the magnitude and direction of the magnetic force $\vec{F}$, acting on electrons in the sensor turn in the variable magnetic field of element dL, using formula (6) with the accuracy to a constant factor. In view of symmetry of the task, the summary magnetic force, acting on the electrons at some point of the sensor turn, is a doubled sum of the forces from a half - turn of the coil. The required induced emf we obtain by projecting the summary magnetic force on the direction of a tangent in a point of the turn in hand and integrating the projection along a contour of the turn, in accordance with (8).

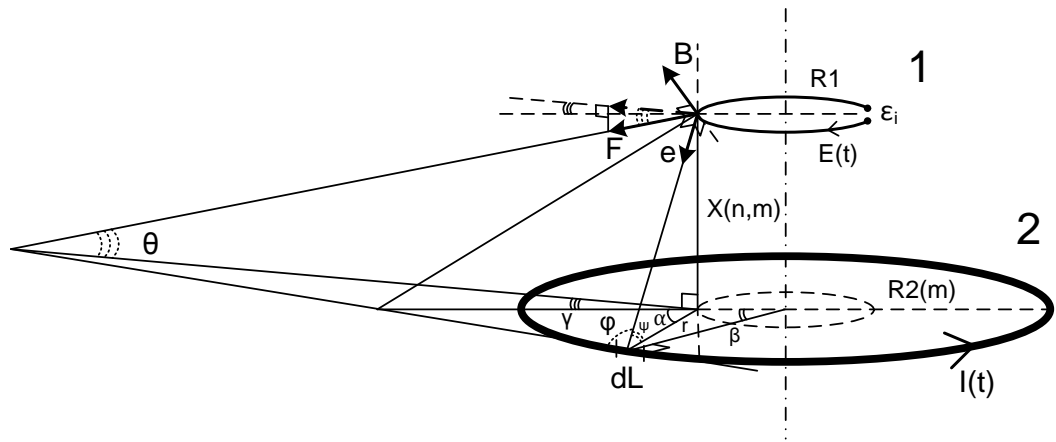

Fig. 11. Explanatory illustration for calculation of the induced emf
in a turn of sensor 1 from a turn of magnetic coil 2.

The calculations are carried out using MathCad 14 software. The expression for absolute value of emf induced in a turn of the sensor by variable magnetic field of the coil is:

$$\varepsilon(k,R1,t) := k \cdot 2\mu_0 \cdot \frac{d}{dt} I(t) \cdot \sum_{m=1}^{5} \sum_{n=1}^{4} \left( R1 \cdot R2(m) \cdot \int_{0}^{\pi} \frac{\alpha'(R1,m,n) - \alpha}{|\alpha'(R1,m,n) - \alpha|} \cdot \frac{\sin(\gamma(R1,\alpha,m,n)) \cdot \sin(\phi(R1,\alpha,m,n)) \cdot \cos(\theta(R1,\alpha,m,n)) \cdot \cos(\psi(R1,\alpha,m,n))^2}{r(R1,\alpha,m)^2} \cdot \frac{d}{d\alpha} \beta(R1,\alpha,m) \, d\alpha \right)$$

(13)

Here k is the desired proportionality factor in the law (8); μ0 is the universal magnetic constant; I(t) is the time dependence of the current, flowing in a winding of the coil; α, β, γ, φ, θ, ψ are the angles- functions (see Fig. 11); α' is the angle, at which the projection of the magnetic force on the tangent to a turn of the sensor at the point with α=0 changes its sign; r is the projection on the plane of the sensor turn of the segment between element *dL* and the point of the turn under consideration. The dependence *I(t)* is obtained by calculation of damped oscillations in the circuit with magnetic coil (see Fig. 2), it takes the form:

$$I(t) := \frac{U0 \cdot e^{\frac{-R \cdot t \cdot 10^{-6}}{2 \cdot L}} \cdot \sin\left(\frac{t \cdot 10^{-6}}{\sqrt{L \cdot C}}\right)}{\sqrt{\frac{L}{C}}} \qquad (14)$$

Here *U0* is the voltage of charged capacitor bank, *R*=20 mOhm is the active resistance of the circuit, *L*=1.9 µH is the inductivity of the coil, *C*= 2850 µF is the capacity of the capacitor bank, *t* is the time in µsec. Inductivity and active resistance are selected so that the shape and time of the dependence correspond to the experimental curves of the magnetic field inside the coil B(t), obtained by Hall sensor. The curve *I(t)* is given in Fig. 12 for the first two pulses. Other functions in (13) are explicit and determined using methods of stereometry. The angle α' is got by finding of zero of the function γ(R1,α,m,n) by α.

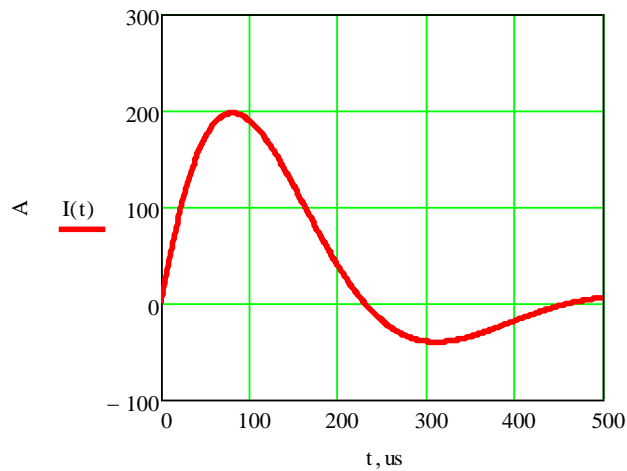

Fig. 12. The calculated time dependence of current in magnetic coil.

The proportionality factor *k* in the expression (13) is determined by comparison of the calculated values of the induced emf with five values measured using the induction sensor with diameter 5.5 mm, and by introduction of a correction. Calculated curves of the emf induced in a winding (10 turns) of the induction sensor are presented in Fig. 13 a, b. The curve of time dependence (a) coincides with the experimental one within the measurement error. The curve of the emf dependence on the turn radius (b) has a square- law shape to a first approximation, that is in accordance with the law (3). The curve discontinuities are caused by imperfection of the search algorithm for the roots of an equation in MathCad. However, shift of the minimum of dependence (b) from the origin of coordinates points to imperfection of the calculation model. One of the reasons of this problem are unaccounted slopes of the coil turns owing to the continuous and irregular winding of the bus-bar. In this connection, a correction is introduced in desired coefficient *k* that halves the previous value of *k*. The correction ensures the agreement between the calculated and experimental values of the induced emf when shifting the minimum of curve (b) to the origin of coordinates. Thus, the desired value of proportionality factor in (8), (13) is: $k = 424 \pm 93 \ m^{-1}$. The error is defined by spread of the experimental values of the induced emf. Dimension of the coefficient is obtained by comparison of the expressions for the induced emf (3) and (8).

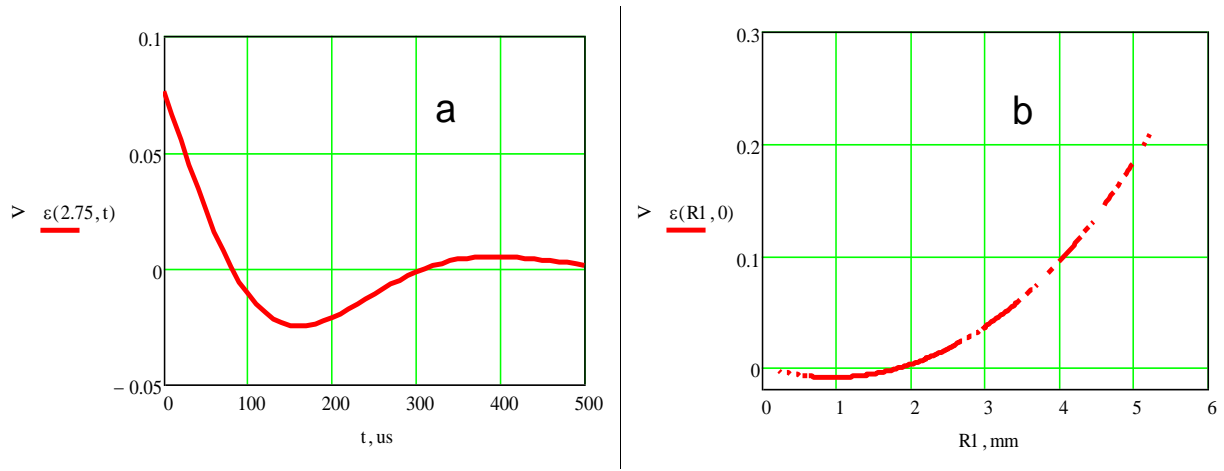

Fig. 13. Calculated curves of the induced EMF in the induction sensor winding. a is the time-dependence when radius of the winding is 2.75 mm; b is the dependence on radius of the winding at initial moment. The voltage of charged capacitor bank is 10 V. The number of turns is 10.

A more exact experiment is needed to specify the proportionality factor in the law (8), with different configurations of the magnetic field source and forms of conductors. Accuracy of calculation of the induced emf will depend on the rate of partition of the field source into the sets of dipoles (elements of the current) as well.

## Conclusion

It has been ascertained that the well- known law of EMI (3) describes the induced emf in conductive turns placed axially inside the magnetic coil. Measurements of the induced emf in wire arcs with different lengths and radii have been carried out. Dependence of the induced emf on the arc length proved to be linear that is in agreement with the law (3) applied to a turn. However, this dependence does not fit the rigorous calculation of the induced emf in an arc made by the law (3). Observed dependence of the induced emf on position of the magnetic field source (coil winding) with regard to the conductor tells about need to take into account the symmetry of influence of magnetic field on the conductor. The shape of the emf curve, induced in a straight wire, is similar to this one for turns and arcs.

The general law of EMI for curvilinear conductor (8) is deduced from the obtained experimental data. In deducing, the magnetic forces acting on a motionless charge ($\bar{v} = 0$) in a variable magnetic field, as well as Lorentz force are considered. In a similar way the expressions for emf (voltage) induced between points of a conductor of arbitrary form can be obtained. Calculation of the induced emf in the sensor winding and comparison with the measured values are carried out to verify the law (8) and to determine the proportionality factor. The obtained dependence of emf on a radius is square- law to a first approximation that fits the law (3). The time dependence of emf coincides with the experimental one within the measurement error. The general law requires further testing for different configurations of the magnetic field relative to conductors.

When a current is induced in a conductor, it is necessary to take into account the decrease of emf due to the self- induction effect and similarity of distribution of the magnetic field near the conductor along the investigated circuit. The induced emf is absent when the distribution is similar and heterogeneities in the circuit are absent. In this case the magnetic forces will synchronously generate induction currents in every part of the conductor. Theoretical estimation of induction current in a conductor with absence of voltages is given above.

Also it is necessary to note, the expressions for the induced emf are obtained under the assumption of high concentration of free charges (elasticity of electron gas). If the concentration is low (semiconductors, dielectrics) the induced emf will be lower significantly. Thus, the induced emf in materials with low electrical conductivity, as well as in ionized conductors should be investigated separately.

Discrimination of the forces acting on a quasi- stationary thermal charge ($\overrightarrow{F_M}$) and on a moving one ($\overrightarrow{F_L}$) from magnetic field is relative in general. The nature of these forces is similar and is connected with interaction of the induced magnetic moment of the charge with the magnetic field. Absence of interaction between free electrons moving along magnetic power lines and the field points to that own magnetic moment (spin) of the free electron is nill. Its magnetic properties are demonstrated only when moving across the magnetic power lines. In this case, the magnetic moment differs from dipole moment and has special character.

Understanding and application of the mechanism of electromagnetic induction may be of use in electrical engineering. For example, the fact that the induced emf depends on position of a source of magnetic field relative to a conductor can be used for development of special induction sensors. It will enable measurement of magnetic induction, as well as determination of spatial structure of the source. It will be of use in magnetography and magnetic tomography.

## Acknowledgements

The author wishes to thank O. A. Petrenko, mechanical engineer of the Chair of General Physics1, for his help in magnetic coils manufacture and Dr. M. V. Balabas for his valuable comments on the work.